\documentclass[review]{elsarticle}

\usepackage{graphicx}
\usepackage{booktabs}
\usepackage{amsmath}
\usepackage{lineno,hyperref}
\usepackage{caption}
\usepackage{subcaption}
\usepackage{setspace}
\usepackage{multicol}
\usepackage{float}
\usepackage[british]{babel}
\usepackage{amsmath,amssymb,bm}

\usepackage[margin=1in]{geometry}

\modulolinenumbers[5]

\begin{document}


\begin{frontmatter}        

\title{The zero-inflated cure rate regression model: Applications to fraud detection in bank loan portfolios}


%
\author{Francisco Louzada\fnref{myfootnote}}
\address{Institute of Mathematical Science and Computing at the University of S{\~a}o Paulo (USP), Brazil} 
\fntext[myfootnote]{Corresponding author: louzada@icmc.usp.br}

\author{Mauro R. de Oliveira Jr.}
\address{Caixa Econ{\^o}mica Federal and Federal University of S{\~a}o Carlos, Brazil}
%

\author{Fernando F. Moreira}
\address{University of Edinburgh Business School, Scotland, UK}
%




\begin{abstract}
                                                                                     
In this paper, we introduce a methodology based on the zero-inflated cure rate model to detect fraudsters in bank loan applications. Our approach enables us to accommodate three different types of loan applicants, i.e., fraudsters, those who are susceptible to default and finally, those who are not susceptible to default. An advantage of our approach is to accommodate zero-inflated times, which is not possible in the standard cure rate model. To illustrate the proposed method, a real dataset of loan survival times is fitted by the zero-inflated Weibull cure rate model. The parameter estimation is reached by maximum likelihood estimation procedure and Monte Carlo simulations are carried out to check its finite sample performance.
\end{abstract}

\begin{keyword}
\texttt {bank loans \sep fraud detection  \sep portfolios \sep survival \sep  zero-inflated \sep Weibull }
\end{keyword}

\end{frontmatter}


\section{Introduction}\label{introd}


In some cases, banks lose contact with customers as soon as their loans are granted and therefore all amount lent is lost. This kind of borrowers are considered fraudsters with, by definition, loan survival time equal to zero. On the other hand, there are customers who no longer honour their loan repayments, but unlike fraudsters, they honour their commitments for a while, until, by private financial reasons, they no more meet their debts with the bank and become a defaulter, however now, with a positive loan survival time which is represented by the elapsed time until the default.

To complete the picture, and ensure the survival of the bank, there are good customers, those who keep up to date with their obligations and, therefore, there are no records of the event of default. So, for the survival of bankers and mainly the maximization of profits, they must seek to maintain high rate of non-defaulted loans, while the rates of fraudsters and defaulters must be low.

Thus, the analysed dataset here is comprised by customers who, in one way or another, have not honoured their contractual obligations with the bank, either by fraud in the application process, or by loss of creditworthiness over time, along with good clients who honour their obligations and have never experienced the event of default. 

According to the aforementioned scenario, and bringing to the statistical terminology, the dataset of study in this paper comprises three different types of non-negative survival times: survival times starting in zero for fraudsters, positive default times for delinquents, and the absence of registration, or censored time, for non-defaulting clients. These considerations delimit the data we will cover in this paper: a set of zeros, positives and unrecorded (censored) banking loan survival times. Such data must be addressed to make a holistic risk management of the loan portfolio, that is, dealing with fraud prevention, delinquency control and ensure the customer loyalty growth.

For using survival analysis techniques, we must consider the modelling outcome of interest be the survival time after loan concession, also mentioned as customer or loan survival time, which is represented by time to occurrence of the event of default. This has been done in different papers, such as \citep{abad2009modelling,banasik1999not,barriga2015non,tong2012mixture}. The reason for the increased use of survival analysis in credit risk over other modelling techniques, besides allowing monitor over time the credit risk of the loan portfolio, is that it can accommodate censored data, which are not supported, for example, in credit scoring techniques based purely on good and bad client classification, see for instance \citep{lessmann2015benchmarking, louzada2006lifetime,stepanova2002survival}.

Notwithstanding, survival analysis deals with non-negative and censored data, however, generally without excess, or even, the presence of zeros. Unlike survival data analysis, in other areas we can observe most commonly the existence of non-negative data with the presence of zeros, sometimes with excess, usually in count data analysis, see for example \citep{barry2002generalized,conceiccao2013zero,lambert1992zero,lord2005poisson,ospina2012general}. Therefore, it is already a commonplace the expression ``zero-inflated data".

Perhaps it is unhelpful, or cruelly insensitive,  if we consider human survival times equal to zero in clinical trials and medical studies. Hence, it might be why, to the best of our knowledge, we have not found yet study that is willing to account for zero-inflated data in the medical specialized literature and that aims to analyse patient survival time.

However, the same sense of respect expended with clinical trials, in some way, does not seem to be the same required when we deal with credit risk events. On the contrary, information about zero-inflated time should be taken into account in credit risk analysis, and must be useful for identifying clients who apply for loans only for the purpose of defrauding the financial institution by, since the beginning, not honouring its obligations under the credit granted. 




\subsection{Preliminary}

In survival analysis, the random variable $T$ of interest is the time span until the occurrence of a certain event, and it is called lifetime. In industry it is usually associated with time to failure of a machine, in the medical area, for example, can be associated with time to recurrence of a disease under treatment. In credit risk, the failure time is related to the time span up to the occurrence of a loan default. Obviously, $T$ is a non-negative variable and, generally, $T$ is treated as a continuous random variable. 

According to \citep{colosimo2006analise,rinne2008weibull}, there are several functions which completely specify the distribution of a random variable, since they are mathematically equivalent functions. We list: the probability density function (PDF), cumulative distribution function (CDF), the complementary cumulative distribution function (CCDF), the hazard rate, the cumulative hazard rate, and the mean residual life function. Inserted in a context of survival lifetime, the complementary cumulative distribution function (CCDF) is also known as survival function.

The downside to consider the standard survival analysis in credit risk is the mathematical fact that the survival function is a proper survival function (goes to zero as time progresses indefinitely). In other words, the survival function, $S(t)=P(T>t)$, holds:

\begin{equation}
	\displaystyle \lim_{t \to \infty}S(t)=0
\end{equation}

In this framework, it is not contemplated the presence of immune subjects to the effects that lead to the occurrence of the concerned event. Returning to the examples in the medical field, there are patients who, once submitted to treatment, recover completely; that is, they are cured. In credit risk, most customers never experience the condition of being delinquent. Considering this, the traditional survival analysis is not suitable for modelling failure time in those cases where there are immunity to the occurrence of failures.

To handle this problem, \citet{berkson1952survival} proposed a simple model that added the fraction of cured $(p>0)$ into the survival function, getting the following survival expression:

\begin{equation}
	S(t)= p + (1-p)S^{*}(t),
\end{equation}

where $S^{*}$ is the survival function of the subject susceptible to failure, and secondly, $p$ is the proportion of subjects immune to failure (cured). This model is called cure rate model, or long-term survival model.

Unlike $S^{*}$, $S$ is an improper survival function,  since it satisfies:

\begin{equation}
	\displaystyle \lim_{t \to \infty}S(t)=p > 0.
\end{equation}

\subsection{Proposal}

We got to the point of presenting the gap and the contribution of our model to the literature on survival analysis. To the best of our knowledge, there is no literature considering a cure rate model that accounts for subjects who have already experimented the event of interest at the beginning of the considered study.

In this sense, and focusing on the portfolio credit risk context, we define the following proportions to be accommodated in the new proposed model:

\begin{itemize}
	\item $\gamma_0$:  the proportion of zero-inflated times, i.e., related to fraudster;
	\item $\gamma_1$:  the proportion of immune to failure, i.e., related to non-defaulters (cured).
\end{itemize}

Thus, we have the following expression for the improper survival function of a dataset comprised by loan survival times:

\begin{equation}
	S(t)= \gamma_1 + (1-\gamma_0-\gamma_1)S^{*}(t), \ \ \ \ \ \ t \geq 0,
	\label{zisf}
\end{equation}

where $S^{*}$ is the survival function related to the $(1-\gamma_0-\gamma_1)$ proportion of subject susceptible to failure, secondly, $\gamma_1$ is the proportion of subjects immune to failure (cured or non-defaulted), and finally, $\gamma_0$ is the proportion of individuals fraudsters (zero-inflated).  This model is called the zero-inflated cure rate model.

The important fact that differentiates the inflated cure version of the standard cure approach, besides the fact of being improper, is expressed in the second of the following satisfied properties: 

\begin{equation}
	\displaystyle \lim_{t \to \infty}S(t)=\gamma_1 > 0.
\end{equation}

\begin{equation}
	\displaystyle S(0)=1-\gamma_0 < 1.
\end{equation}

Note that, if $\gamma_0=0$, i.e., without the excess of zeros, we have the cure rate model of \citet{berkson1952survival}.

\begin{figure}[htbp]
	\centering
		\includegraphics[scale=0.15]{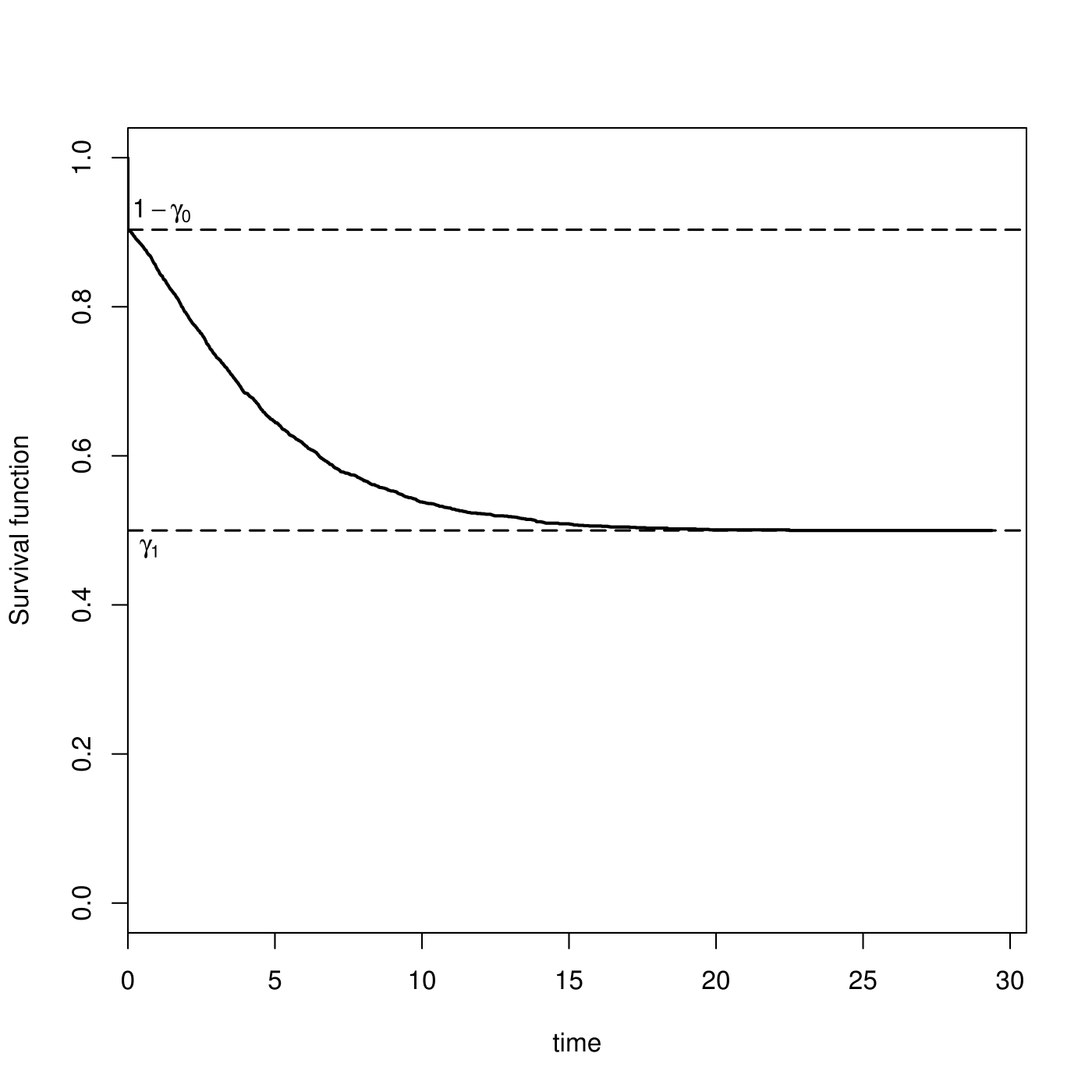}
	\caption{Survival function of the zero-inflated cure rate model}
	\label{fig:fig1_LD_ZAW}
\end{figure}


Here, we justify the need for the zero-inflated cure model based on a credit risk setting. To exemplify the application of the proposed approach, we analyse a portfolio of loans made available by a large Brazilian commercial bank, in order to check whether an applicant is a fraudster and, if not, how and when she or he will (or will not) become a defaulter. This is why we propose a methodology based on an enhanced cure rate approach, for the purposes of dealing with the problem of assessing the propensity to fraud in bank loan applications.




The paper is organized as follows. In Section \ref{model}, we formulate the model and present the approach for parameter estimation. A study based on Monte Carlo simulations with a variety of parameters is presented in Section \ref{sim}. An application to a real data set of a Brazilian bank loan portfolio is presented in Section \ref{apl}. Some general remarks are presented in Section \ref{con}.

\section{Model specification}\label{model}

In what follows, we consider the zero-inflated cure rate model as defined in equation \ref{zisf}. The associated (improper) cumulative distribution function (CDF) and  failure density function (PDF) are given by:

\begin{equation}
	F(t)= \gamma_0 + (1-\gamma_0-\gamma_1)F^{*}(t), \ \ \ \ \ \ t \geq 0,
	\label{zicdf}
\end{equation}

\begin{equation}
f(t) = \left\{
\begin{array}{lcl}
\gamma_0, & \mbox{if} & t = 0,\\
(1-\gamma_0-\gamma_1)f^{*}(t), & \mbox{if} & 0<t, \\
\end{array}
\right.
\label{zipdf}
\end{equation}


where the parameters $\gamma_0$ and $\gamma_1$ are as defined in equation \ref{zisf}, $F^{*}$ and $f^{*}$ are, respectively, the cumulative distribution function and probability density function underpinning the $(1-\gamma_0-\gamma_1)$ proportion of subject susceptible to failure.

Note that, the improper CDF of the zero-inflated cure rate model, $F(t)$, has the advantageous property of accommodating the excess of zeros, $\gamma_0$, since it satisfies: $F(0)=\gamma_0$. Moreover, it does not undermine the fraction of cured, $\gamma_1$, since it also satisfies: $\displaystyle \lim_{t \to \infty}F(t)=1-\gamma_1$.

\subsection{The zero-inflated Weibull cure rate model}

In this paper, we associate the Weibull distribution as the probability density function for the subject susceptible to failure. We choose the Weibull function since it has been widely used to model survival data, and also has served as motivation for the proposal of various types of generalizations, see for example \citep{cancho2013long,cooner2007flexible,ortega2012negative,rinne2008weibull}.

Then, let the Weibull (W) distribution represents the random behaviour of the non negative random variable $T^*$, which denotes the time-to-default for the susceptible subjects to the event of default. The CDF of the W distribution is given by

\begin{equation}
	F_w^* (t) = 1 - e^{- \left( \frac{t}{\lambda} \right)^\alpha}, t \geq 0,
	\label{wcdf}
\end{equation}

where $\alpha > 0$ and $\lambda > 0 $ are, respectively, the shape and scale parameters. The PDF of the W distribution is obtained from the equation \ref{wcdf} as

\begin{equation}
	f_w^* (t) = \frac{\alpha}{\lambda} {\left( \frac{t}{\lambda} \right)}^{\alpha-1} e^{\left( - \frac{t}{\lambda} \right)^\alpha}, t \geq 0.
	\label{wpdf}
\end{equation}

The zero-inflated cure rate model proposes to distinguish between three sub-populations of loan applicants based on its loan survival times and its susceptibility to the event of fraud or default: a segment of those who will not honour any instalment of the loan, i.e., fraudsters, a segment for those are susceptible to default and will eventually default, i.e., defaulters, and finally, a segment for those who are not susceptible to default and will not experience the event of default during the loan term, i.e., cured or non-defaulters or long-term survivors. 

Consequently, consider the indicator function $I(T_i>0)$, where $I(T_i>0)=1$ if $T_i>0$, and $I(T_i>0)=0$ otherwise. This indicator function divides the portfolio into two types of customers, fraudsters and non-fraudsters. Following the settings as done in \citep{tong2012mixture}, let $M$ be a binary random variable defined for the susceptibility to default event, with $M= 0$ denoting that the account is non-susceptible and so a long-term survivor, while $M= 1$ states that the account is susceptible and will default at some time point, or it may be censored in the dataset. Still following the notations given in \citep{tong2012mixture},  let us define a censoring indicator $\delta$, where $\delta=1$ indicates non-censored lifetimes and $\delta=0$ indicates censored lifetimes. 


There are two possibilities for the customer who is not a fraudster and has not defaulted during the follow-up of the loan portfolio. Information about the default time is right censored, that is, the client will probably become a defaulter if given enough time, or the client is really a good payer and will never default, regardless of the monitoring period term. There are then four possible states of the data as follows:

\begin{itemize}
\item  $I(T>0)=0$: the client is a fraudster;
\item  $I(T>0)=1$, $\delta=1$ and $M = 1$: non-fraudster, non-censored, susceptible, hence the client has defaulted;
\item  $I(T>0)=1$, $\delta=0$ and $M = 1$: non-fraudster, censored, susceptible, hence the client has not defaulted but would eventually default in future;
\item $I(T>0)=1$, $\delta=0$ and $M = 0$: non-fraudster, censored, non-susceptible, hence cured or long-term survivor.
\end{itemize}

\subsection{Likelihood function}

For the likelihood contribution of each client $i$, we should pay attention to fact that there are different sub-group of clients. Therefore, the likelihood contribution of a client $i$ for the zero-inflated cure rate model, obtained from \ref{zisf} and all considerations we have done, must assume three different values:

\begin{equation}
\left\{
\begin{array}{lcl}
\delta_0,& \mbox{if} & i \mbox{ is a fraudster},\\
(1 - \gamma_0 - \gamma_1)f_w^* (t_i), & \mbox{if} & i \mbox{ is not censored}  \\
\gamma_1 + (1-\gamma_0-\gamma_1)S^{*}(t_i), & \mbox{if} & i \mbox{ is censored},
\end{array}
\right.
\label{ziwpdf}
\end{equation}

Let the data take the form $\mathcal{D} = \left\{{t_{i},\delta_{i}}\right\}$, where $\delta_i=1$ if $t_i$ is a observable time to default, and  $\delta_i=0$ if it is right censored, for $i = 1,2,\cdots n.$ Let $(\alpha,\lambda)$ denote the parameter vector of the W distribution and, finally, $(\gamma_0,\gamma_1)$ the parameters associated, respectively, with the proportion of fraudsters (inflation of zeros) and the proportion of long-term survivors (cure rate).

The likelihood function of the inflated-zero Weibull cure rate model, with a four-parameter vector $\vartheta = (\alpha,\lambda,\gamma_0,\gamma_1 )$, is based on a sample of $n$ observations, $\mathcal{D} = \left\{{t_{i},\delta_{i}}\right\}$. Then, we can write the likelihood function under non-informative censoring as 

\begin{equation}
L(\vartheta ; \mathcal{D}) \propto  \prod_{i=1}^n \left\{ (1-I(t_i>0)) \gamma_0 + I(t_i>0) {\left[(1 - \gamma_0 - \gamma_1)f_w^* (t_i)\right]}^{\delta_i} {\left[\gamma_1 + (1-\gamma_0-\gamma_1)S^{*}(t_i)\right]}^{1-\delta_i}  \right\} 
\label{like}
\end{equation}

\subsection{The regression version}


Here, we introduce an approach to accommodate covariates in a regression setting. Therefore, we propose to relate the set of four parameters, $(\gamma_0,\gamma_1,\alpha,\lambda)$, respectively, the proportion of zeros, the proportion of cured, the shape and the scale parameter of the W distribution, with a set of four-covariate vectors, $\mathbf x_1$, $\mathbf x_2$,$\mathbf x_3$ and $\mathbf x_4$, respectively. These covariate vectors, as occurs in practice, may be the same, i.e., $\mathbf x_1=\mathbf x_2=\mathbf x_3=\mathbf x_4$.

Following the setting made in \citep{pereira2013regression}, p. 128, the regression version of the zero-inflated Weibull cure rate model is defined by \ref{zisf} up to \ref{ziwpdf}, and by the following components, known as link functions:

\begin{equation}
 \left\{
\begin{array}{llll}
(\gamma_{0i},\gamma_{1i}) &=& \left(\frac{e^{\mathbf x^\top_{1i}\beta_1}}{1+e^{\mathbf x^\top_{1i}\beta_1}+e^{\mathbf x^\top_{2i}\beta_2}},\frac{e^{\mathbf x^\top_{2i}\beta_2}}{1+e^{\mathbf x^\top_{1i}\beta_1}+e^{\mathbf x^\top_{2i}\beta_2}}\right),\\
\alpha_{i}               &=& e^{\mathbf x^\top_{3i}\beta_3}, \\
\lambda_{i}               &=& e^{\mathbf x^\top_{4i}\beta_4}, \\
\end{array}
\right.
\label{ziwcmR}
\end{equation}

where $\beta_j$'s	are four vectors of regression coefficients to be estimated.

\subsection{Parameter estimation}

Parameter estimation is performed by straightforward use of maximum likelihood estimation (MLE), where, as we will see, its simple application is supported by our simulation studies.

The maximum likelihood estimates $\hat{\vartheta}$, regarding the parameter vector $\vartheta$, are obtained through maximization of $L(\vartheta;\mathcal{D})$ or $\ell(\vartheta;\mathcal{D})=\log\{L(\vartheta;\mathcal{D})\}$. Under suitable regularity conditions, the asymptotic distribution of the maximum likelihood estimates (MLEs), $\hat{\vartheta}$, is a multivariate normal with mean vector $\vartheta$ and covariance matrix, which can be estimated by $\{-\partial^2\ell(\vartheta)/\partial{\vartheta}\partial{\vartheta}^T\}^{-1}$, evaluated at $\vartheta=\hat{\vartheta}$, where the required second derivatives are computed numerically, see \citep{migon2014statistical}. There are many software and routines available for numerical maximization. We choose the method ``BFGS" for maximizing, see details in \citep{manualR}, which comes within the \textbf{R} routine \texttt{optim}.

\section{Simulation Studies}\label{sim}

We proceed a parameter estimation based on a maximum likelihood principle and use the R routine optim() for that. In order to assess the performance of the maximum likelihood estimates with respect to sample size, i.e., to ensure that the root mean square errors, the relative bias and standard deviations decrease as sample size increases, we perform Monte Carlo simulations, where the sample size varies as $n=100, 250, 500, 1000, 2000$ and each sample is replicated 100 times. The description of sample generation, details of the simulated time distribution, and results obtained regarding the proposed model are described below. 

Three simulation studies are performed for the proposed zero-inflated Weibull cure rate regression model, which was introduced by the improper survival function in \ref{zisf} and the link functions in \ref{ziwcmR}. For this purpose of simulation, we let $x$ be a random variable that represents a consumer characteristic. Hence, the configuration of parameters on a single variable $x$ is replaced by the following expression:

\begin{equation}
 \left\{
\begin{array}{llll}
(\gamma_{0i},\gamma_{1i}) &=& \left(\frac{e^{\mathbf \beta_{10} + x_i\beta_{11}}}{1+e^{\mathbf \beta_{10} + x_i\beta_{11}}+e^{\mathbf \beta_{20} + x_i\beta_{21}}},\frac{e^{\mathbf \beta_{20} + x_i\beta_{21}}}{1+e^{\mathbf \beta_{10} + x_i\beta_{11}}+e^{\mathbf \beta_{20} + x_i\beta_{21}}}\right),\\
\alpha_{i}               &=& e^{\mathbf \beta_{30} + x_i\beta_{31}}, \\
\lambda_{i}               &=& e^{\mathbf \beta_{40} + x_i\beta_{41}}, \\
\end{array}
\right.
\label{apliziwcmR}
\end{equation}

\subsection{Simulation algorithm}

Suppose that the time of occurrence of an event of interest has the improper cumulative distribution function $F(t)$ given by \ref{zicdf}. Actually, we want to simulate a random sample of size n containing real times, with a proportion $\gamma_0$ of zero-inflated times, censored times and a cure fraction of $\gamma_1$, and with a proportion of failures times drawn from a Weibull distribution with $\alpha$ and $\lambda$ parameters.
 
\begin{enumerate}
\item determine $\beta_{10}$ and $\beta_{11}$ related to the value of the desired proportion of zero-inflated times, $\gamma_0$; determine $\beta_{20}$ and $\beta_{21}$ related to the value of the desired cure fraction, $\gamma_1$; finally, set the Weibull parameters $\beta_{30}$ and $\beta_{31}$, related to $\alpha$, as well as, $\beta_{40}$ and $\beta_{41}$ related to $\lambda$;
\item drawn $x_i$ from $x$ and calculate $\gamma_{0i}$, $\gamma_{1i}$, $\alpha_i$ and $\lambda_i$ as in \ref{apliziwcmR};
\item generate $u_i$ from a uniform distribution U(0,1);
\item if $u_i \leq \gamma_{0i}$, set $s_i=0$;
\item if $u_i > 1- \gamma_{1i}$, set $s_i=\infty$;
\item if $\gamma_{0i} < u_i \leq 1-\gamma_{1i}$, generate $v_i$ from a uniform distribution U$(\gamma_{0i},1-\gamma_{1i})$ and take $s_i$ as the root of $F(t)-v_i=0$;
\item generate $w_i$ from a  uniform U$(0, max(s_i))$, considering only finites $s_i$;
\item calculate $t_i =min(s_i,w_i)$, if $t_i < w_i$, set $\delta_i=1$,  otherwise, set $\delta_i=0$.
\item Repeat as necessary from step 2 until you get the desired amount of sample $(t_i,\delta_i)$.
\end{enumerate}

Note that the censoring distribution chosen is a uniform distribution with limited range in order to keep the censoring rates reasonable, see \citep{rocha2015new}, p.12.

\subsection{Parameter scenarios}

Considering the settings parameters established in the regression model defined in \ref{apliziwcmR}, we set three different scenarios of parameters for the simulation studies here performed. Playing the role of covariate,  we assume $x$ as a binary covariate with values drawn from a Bernoulli distribution with parameter 0.5.

For scenario 1, $\beta_{10}$ assumes -3 and $\beta_{11}$ assumes 1. Given the average value of $x$ is 0.5, we have that $\gamma_0$ assumes on average a value of 0.0697.  $\beta_{20}$ assumes -2.5 and $\beta_{21}$ assumes 0.3, and then,  $\gamma_1$ assumes on average a value of 0.0809. Compared to the other scenarios, scenario 1 has the characteristic of having a \textbf{low rate of fraudsters and cured}, respectively, 6,97\% and 8,09\%.
Regarding to the Weibull parameters,  $\beta_{30}$ assumes 0.5, $\beta_{31}$ assumes 0.5,  $\beta_{40}$ assumes 1.5 and $\beta_{41}$ assumes 2. This implies that the Weibull average parameters $\alpha$ and $\lambda$  are, respectively, equal to 2.117 and 12.182. In this scenario 1, the mean and standard deviation  of the defaulted time are respectively equal to 10.789 and 5.358.

For scenario 2, $\beta_{10}$ assumes -2 and $\beta_{11}$ assumes 2. Given the average value of $x$ is 0.5, we have that $\gamma_0$ assumes on average a value of 0.1999.  $\beta_{20}$ assumes -1.5 and $\beta_{21}$ assumes 1.5, and then,  $\gamma_1$ assumes on average a value of 0.256. Compared to the other scenarios, scenario 2 has the characteristic of having a \textbf{moderate rate of fraudsters and cured}, respectively, 19.99\% and 25.6\%.
Regarding to the Weibull parameters,  $\beta_{30}$ assumes -0.5, $\beta_{31}$ assumes 1.5,  $\beta_{40}$ assumes -0.5 and $\beta_{41}$ assumes 3. This implies that the Weibull average parameters $\alpha$ and $\lambda$  are, respectively, equal to 1.284 and 2.718. In this scenario 2, the mean and standard deviation  of the defaulted time are respectively equal to 2.516 and 1.975.

Finally, for scenario 3, $\beta_{10}$ assumes -0.5 and $\beta_{11}$ assumes 0.75. Given the average value of $x$ is 0.5, we have that $\gamma_0$ assumes on average a value of 0.2469.  $\beta_{20}$ assumes -0.35 and $\beta_{21}$ assumes 1.75, and then,  $\gamma_1$ assumes on average a value of 0.4731. Compared to the other scenarios, scenario 3 has the characteristic of having a \textbf{high rate of fraudsters and cured}, respectively, 24.69\% and 47.31\%.
Regarding to the Weibull parameters,  $\beta_{30}$ assumes 1, $\beta_{31}$ assumes 2,  $\beta_{40}$ assumes 1.25 and $\beta_{41}$ assumes 3.5. This implies that the Weibull average parameters $\alpha$ and $\lambda$  are, respectively, equal to 1 and 20.085. In this scenario 3, the mean and standard deviation  of the defaulted time are both equal to 20.085.

\subsection{Results of Monte Carlo simulations}

\begin{figure}[H]
	\centering
		\includegraphics[scale=0.3375]{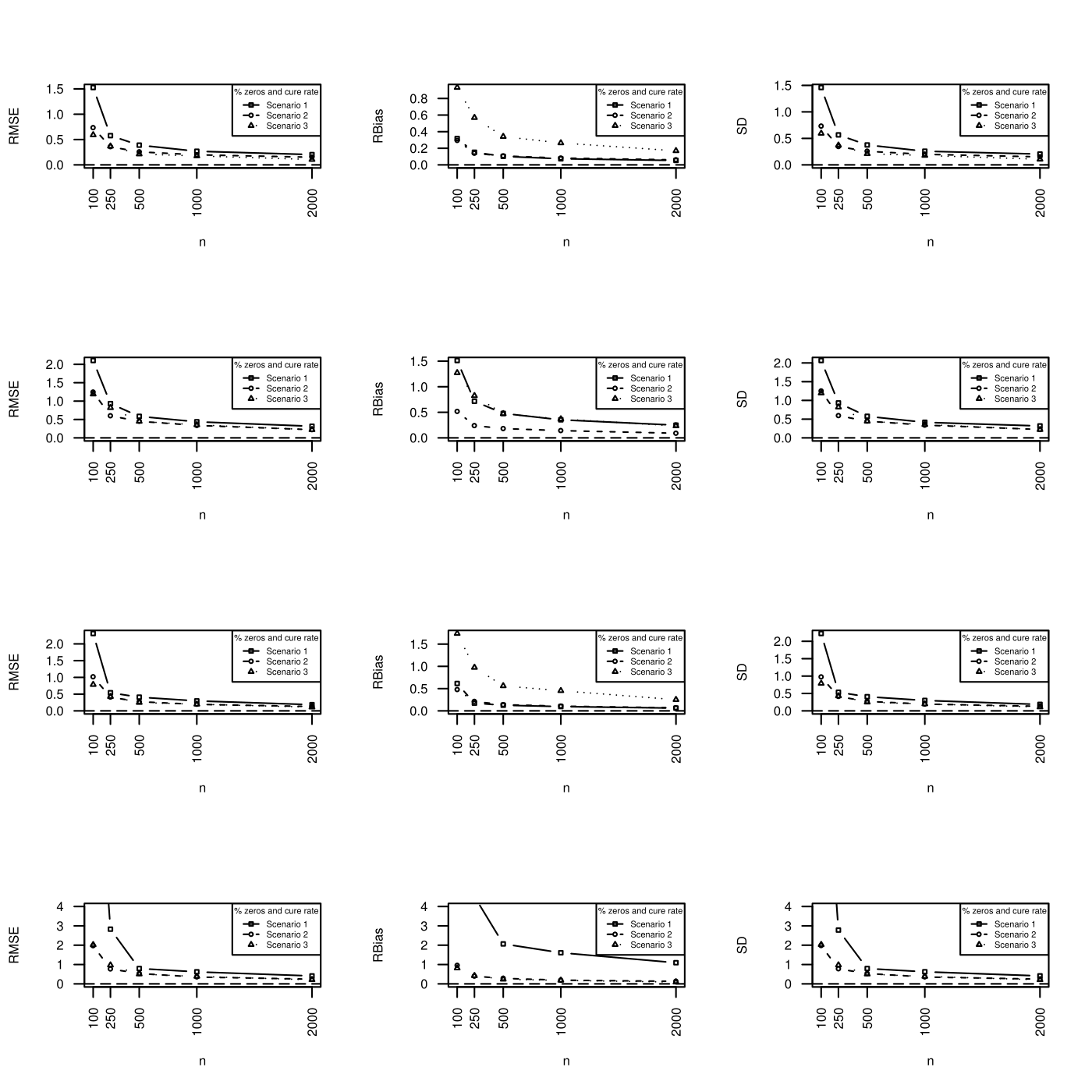}
	\caption{Root mean square error, relative bias and standard deviation of the estimated regression parameters $\beta_{10},\beta_{12},\beta_{20},\beta_{21}$, respectively related to $\gamma_0$ and $\gamma_1$,  of the simulated data from the zero-inflated Weibull cure rate regression model, obtained from Monte Carlo simulations with 100 replications, increasing sample size $(n)$.}
	\label{fig:sim_all_123_LDZAW2}
\end{figure}

\begin{figure}[H]
	\centering
		\includegraphics[scale=0.3375]{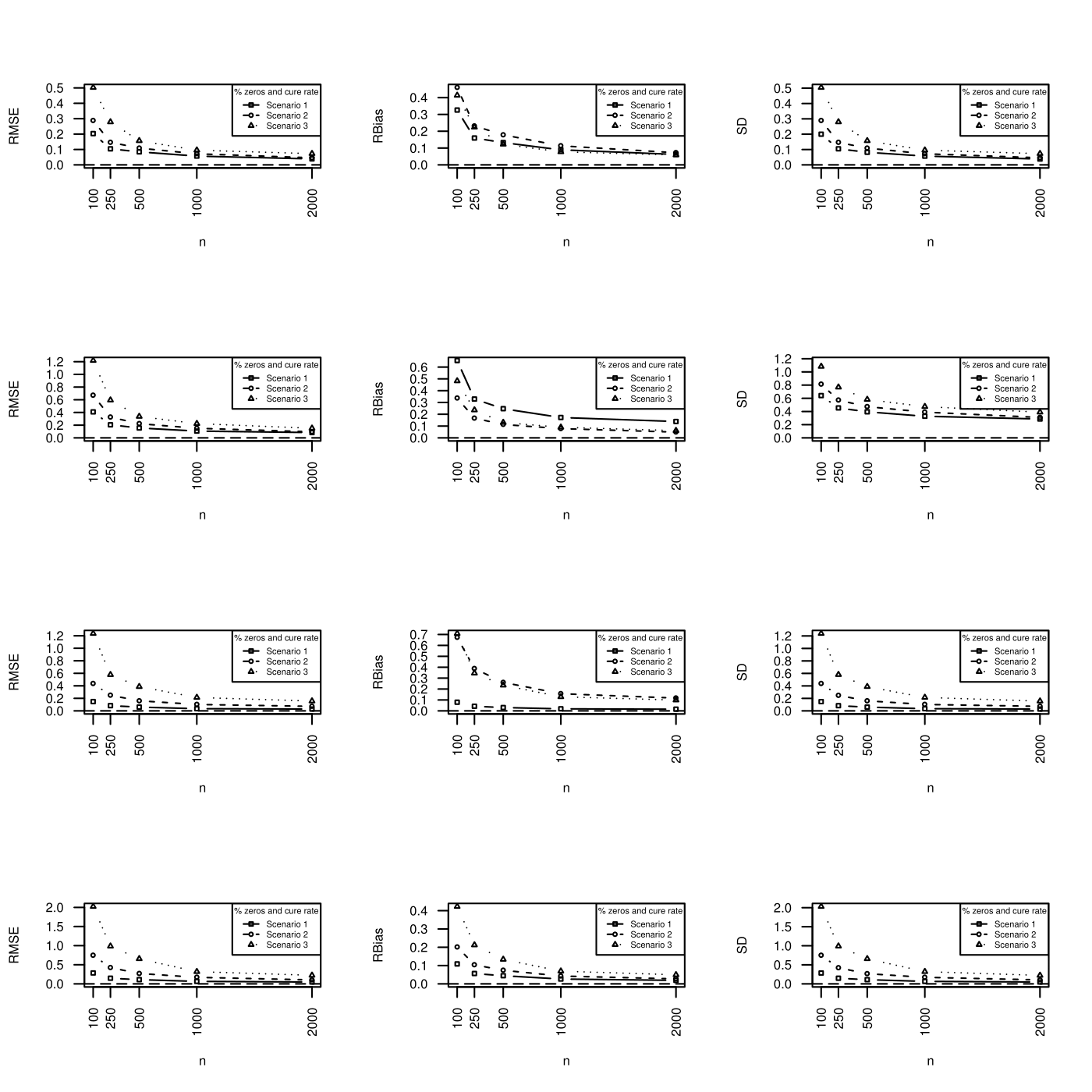}
	\caption{Root mean square error, relative bias and standard deviation of the estimated regression parameters $\beta_{30},\beta_{32},\beta_{40},\beta_{41}$, respectively related to $\alpha$ and $\lambda$,  of the simulated data from the zero-inflated Weibull cure rate regression model, obtained from Monte Carlo simulations with 100 replications, increasing sample size $(n)$.}
	
	\label{fig:sim_all_123_LDZAW1}
\end{figure}

Figures \ref{fig:sim_all_123_LDZAW2} and \ref{fig:sim_all_123_LDZAW1}, among with tables \ref{LT1}, \ref{LT3} and \ref{LT5}, describe the simulation results for three different scenario of parameters as described above. Note that relative bias are showed only in the figures. The parameter values were selected in order to assess the estimation method performance under different scales and shapes for the time-to-default distribution, and also under a composition of different proportions of fraudsters and defaulters.

The following can be observed from the figures and tables:

\begin{itemize} 
\item in general, the maximum likelihood estimation in average, MLEA, is closer to the parameters set in the simulated sampling, as sample size increases; 
\item in general, the root mean square error, relative biases and standard deviations decrease as sample size increases;
\item comparing figures \ref{fig:sim_all_123_LDZAW2} and \ref{fig:sim_all_123_LDZAW1}, as expected, we can see that in the scenarios with the greatest presence of fraudsters and cured, i.e., scenarios 2 and 3, the measures of RMSE an SD of the regression parameters related to $\gamma_0$ and $\gamma_1$, see figure \ref{fig:sim_all_123_LDZAW2}, decrease faster than the respective measures related to the estimated regression parameters of the population susceptible to default, i.e., $\alpha$ and $\lambda$, see in figure \ref{fig:sim_all_123_LDZAW1}.
\end{itemize}

\begin{table}[H]
\centering
\caption{Maximum likelihood estimation in average, root mean square error, and standard deviation of the estimated parameters of the simulated data from the zero-inflated Weibull cure rate regression model, obtained from Monte Carlo simulations with 100 replications, increasing sample size $(n)$ and $29,08\%$ in average of censored data (CD).}

\centering
\smallskip\noindent
\resizebox{\linewidth}{!}{

\begin{tabular}{|c|ccc|ccc|ccc|ccc|}
\toprule

&\multicolumn{6}{c|}{$\overline{\gamma_0}=0.0697$} & \multicolumn{6}{|c|}{$\overline{\gamma_1}=0.0809$} \\ 

\cline{2-13}

&\multicolumn{3}{c|}{$\beta_{10}=-3.0$}&\multicolumn{3}{|c}{$\beta_{11}=1.0$}& \multicolumn{3}{|c|}{$\beta_{20}=-2.5$}&\multicolumn{3}{c|}{$\beta_{21}=0.30$} \\ 

\cline{2-13}

$n$ (CD) & \multicolumn{1}{c}{MLEA} & \multicolumn{1}{c}{RMSE} & \multicolumn{1}{c|}{SD} & \multicolumn{1}{c}{MLEA} & \multicolumn{1}{c}{RMSE} & \multicolumn{1}{c|}{SD} & \multicolumn{1}{c}{MLEA} & \multicolumn{1}{c}{RMSE}  &\multicolumn{1}{c|}{SD} & \multicolumn{1}{c}{MLEA} & \multicolumn{1}{c}{RMSE} & \multicolumn{1}{c|}{SD} \\
\hline
        
100 (33.3\%)	&		-3.4696	&	1.5270	&	1.4603	&	1.4522	&	2.1039	&	2.0651	&	-3.1656	&	2.3107	&	2.2239	&	-2.1177	&	15.0721	&	14.9518	\\
250 (30.1\%)	&		-3.1259	&	0.5768	&	0.5657	&	1.0860	&	0.9316	&	0.9323	&	-2.3953	&	0.5418	&	0.5343	&	-0.2862	&	2.8293	&	2.7819	\\
500 (28.7\%)	&		-3.1069	&	0.3892	&	0.3761	&	1.1230	&	0.5851	&	0.5749	&	-2.5604	&	0.4097	&	0.4072	&	0.3885	&	0.7938	&	0.7929	\\
1000 (27.0\%)	&		-2.9208	&	0.2688	&	0.2582	&	0.8566	&	0.4375	&	0.4154	&	-2.5040	&	0.2981	&	0.2995	&	0.2937	&	0.6220	&	0.6251	\\
2000 (26.3\%)	&		-3.0103	&	0.2029	&	0.2037	&	0.9991	&	0.3164	&	0.3180	&	-2.4843	&	0.1855	&	0.1858	&	0.2608	&	0.4118	&	0.4120	\\

\end{tabular}\label{LT1}
}
\centering
\smallskip\noindent
\resizebox{\linewidth}{!}{
\begin{tabular}{|c|ccc|ccc|ccc|ccc|}
\toprule
&\multicolumn{6}{c|}{$\overline{\alpha}=2.117$} & \multicolumn{6}{c|}{$\overline{\lambda}=12.182$} \\ 
\cline{2-13}
&\multicolumn{3}{c|}{$\beta_{30}=0.50$}&\multicolumn{3}{c|}{$\beta_{31}=0.50$}& \multicolumn{3}{|c|}{$\beta_{40}=1.50$}&\multicolumn{3}{|c|}{$\beta_{41}=2.0$} \\ 
\cline{2-13}
$n$ (CD) & \multicolumn{1}{c}{MLEA} & \multicolumn{1}{c}{RMSE} & \multicolumn{1}{c|}{SD} & \multicolumn{1}{c}{MLEA} & \multicolumn{1}{c}{RMSE} & \multicolumn{1}{c|}{SD} & \multicolumn{1}{c}{MLEA} & \multicolumn{1}{c}{RMSE}  &\multicolumn{1}{c|}{SD} & \multicolumn{1}{c}{MLEA} & \multicolumn{1}{c}{RMSE} & \multicolumn{1}{c|}{SD} \\
\hline
       
100 (33.3\%)	&		0.5407	&	0.2032	&	0.2001	&	0.4821	&	0.4092	&	0.4109	&	1.5151	&	0.1482	&	0.1481	&	1.9714	&	0.2852	&	0.2851	\\
250 (30.1\%)	&		0.5035	&	0.1042	&	0.1047	&	0.5175	&	0.2058	&	0.2060	&	1.4869	&	0.0836	&	0.0830	&	2.0142	&	0.1481	&	0.1481	\\
500 (28.7\%)	&		0.5226	&	0.0847	&	0.0821	&	0.4877	&	0.1558	&	0.1561	&	1.5052	&	0.0602	&	0.0602	&	1.9927	&	0.1105	&	0.1108	\\
1000 (27.0\%)	&		0.5049	&	0.0571	&	0.0572	&	0.4969	&	0.1072	&	0.1077	&	1.4974	&	0.0352	&	0.0353	&	2.0037	&	0.0655	&	0.0657	\\
2000 (26.3\%)	&		0.5079	&	0.0389	&	0.0383	&	0.4828	&	0.0834	&	0.0821	&	1.5019	&	0.0292	&	0.0293	&	1.9972	&	0.0487	&	0.0489	\\

\bottomrule
\end{tabular}\label{LT2}
}

\end{table}

\begin{table}[H]
\centering
\caption{Maximum likelihood estimation in average, root mean square error, and standard deviation of the estimated parameters of the simulated data from the zero-inflated Weibull cure rate regression model, obtained from Monte Carlo simulations with 100 replications, increasing sample size $(n)$ and $34,91\%$ in average of censored data (CD).}

\centering
\smallskip\noindent
\resizebox{\linewidth}{!}{

\begin{tabular}{|c|ccc|ccc|ccc|ccc|}

\toprule

&\multicolumn{6}{c|}{$\overline{\gamma_0}=0.1999$} & \multicolumn{6}{c|}{$\overline{\gamma_1}=0.2560$} \\ 

\cline{2-13}

&\multicolumn{3}{c|}{$\beta_{10}=-2.0$}&\multicolumn{3}{c|}{$\beta_{11}=2.0$}& \multicolumn{3}{c|}{$\beta_{20}=-1.5$}&\multicolumn{3}{c|}{$\beta_{21}=1.5$} \\ 

\cline{2-13}

$n$ (CD) & \multicolumn{1}{c}{MLEA} & \multicolumn{1}{c}{RMSE} & \multicolumn{1}{c|}{SD} & \multicolumn{1}{c}{MLEA} & \multicolumn{1}{c}{RMSE} & \multicolumn{1}{c|}{SD} & \multicolumn{1}{c}{MLEA} & \multicolumn{1}{c}{RMSE}  &\multicolumn{1}{c|}{SD} & \multicolumn{1}{c}{MLEA} & \multicolumn{1}{c}{RMSE} & \multicolumn{1}{c|}{SD} \\
\hline
        
100 (36.7\%)	&	-2.1023	&	0.7333	&	0.7298	&	2.0543	&	1.2451	&	1.2502	&	-1.8035	&	1.0157	&	0.9741	&	1.6158	&	1.9738	&	1.9804	\\
250 (35.7\%)	&	-2.0956	&	0.3572	&	0.3459	&	2.1190	&	0.6007	&	0.5917	&	-1.5116	&	0.4105	&	0.4124	&	1.4335	&	0.7732	&	0.7742	\\
500 (34.2\%)	&	-1.9717	&	0.2565	&	0.2562	&	1.9264	&	0.4499	&	0.4461	&	-1.5283	&	0.2730	&	0.2729	&	1.4681	&	0.5360	&	0.5377	\\
1000 (34.4\%)	&	-2.0059	&	0.1990	&	0.1999	&	2.0092	&	0.3389	&	0.3405	&	-1.5032	&	0.1950	&	0.1959	&	1.5002	&	0.3545	&	0.3562	\\
2000 (33.6\%)	&	-1.9986	&	0.1510	&	0.1518	&	1.9990	&	0.2232	&	0.2243	&	-1.5066	&	0.1312	&	0.1317	&	1.5075	&	0.2440	&	0.2451	\\

\end{tabular}\label{LT3}
}
\centering
\smallskip\noindent
\resizebox{\linewidth}{!}{
\begin{tabular}{|c|ccc|ccc|ccc|ccc|}
\toprule
&\multicolumn{6}{c|}{$\overline{\alpha}=1.284$} & \multicolumn{6}{c|}{$\overline{\lambda}=2.718$} \\ 
\cline{2-13}
&\multicolumn{3}{c|}{$\beta_{30}=-0.50$}&\multicolumn{3}{c|}{$\beta_{31}=1.50$}& \multicolumn{3}{c|}{$\beta_{40}=-0.50$}&\multicolumn{3}{c|}{$\beta_{41}=3.0$} \\ 
\cline{2-13}
$n$ (CD) & \multicolumn{1}{c}{MLEA} & \multicolumn{1}{c}{RMSE} & \multicolumn{1}{c|}{SD} & \multicolumn{1}{c}{MLEA} & \multicolumn{1}{c}{RMSE} & \multicolumn{1}{c|}{SD} & \multicolumn{1}{c}{MLEA} & \multicolumn{1}{c}{RMSE}  &\multicolumn{1}{c|}{SD} & \multicolumn{1}{c}{MLEA} & \multicolumn{1}{c}{RMSE} & \multicolumn{1}{c|}{SD} \\
\hline
       
100 (36.7\%)	&	-0.5060	&	0.2881	&	0.2895	&	1.6246	&	0.6711	&	0.6627	&	-0.4989	&	0.4360	&	0.4382	&	2.9968	&	0.7489	&	0.7526	\\
250 (35.7\%)	&	-0.4814	&	0.1469	&	0.1464	&	1.5046	&	0.3270	&	0.3286	&	-0.5293	&	0.2493	&	0.2488	&	3.0585	&	0.4250	&	0.4231	\\
500 (34.2\%)	&	-0.4932	&	0.1090	&	0.1094	&	1.4886	&	0.2234	&	0.2243	&	-0.5210	&	0.1611	&	0.1605	&	3.0574	&	0.2726	&	0.2678	\\
1000 (34.4\%)	&	-0.4964	&	0.0719	&	0.0722	&	1.5030	&	0.1515	&	0.1522	&	-0.4968	&	0.1026	&	0.1030	&	3.0019	&	0.1710	&	0.1719	\\
2000 (33.6\%)	&	-0.5036	&	0.0463	&	0.0463	&	1.5105	&	0.0961	&	0.0960	&	-0.4951	&	0.0729	&	0.0731	&	2.9935	&	0.1047	&	0.1050	\\

\bottomrule
\end{tabular}\label{LT4}
}
\end{table}

\begin{table}[H]
\centering
\caption{Maximum likelihood estimation in average, root mean square error, and standard deviation of the estimated parameters of the simulated data from the zero-inflated Weibull cure rate regression model, obtained from Monte Carlo simulations with 100 replications, increasing sample size $(n)$ and $51,96\%$ in average of censored data (CD).}

\centering
\smallskip\noindent
\resizebox{\linewidth}{!}{

\begin{tabular}{|c|ccc|ccc|ccc|ccc|}

\toprule

&\multicolumn{6}{c|}{$\overline{\gamma_0}=0.2469$} & \multicolumn{6}{c|}{$\overline{\gamma_1}=0.4731$} \\ 

\cline{2-13}

&\multicolumn{3}{c|}{$\beta_{10}=-0.50$}&\multicolumn{3}{c|}{$\beta_{11}=0.75$}& \multicolumn{3}{c|}{$\beta_{20}=-0.35$}&\multicolumn{3}{c|}{$\beta_{21}=1.75$} \\ 

\cline{2-13}

$n$ (CD) & \multicolumn{1}{c}{MLEA} & \multicolumn{1}{c}{RMSE} & \multicolumn{1}{c|}{SD} & \multicolumn{1}{c}{MLEA} & \multicolumn{1}{c}{RMSE} & \multicolumn{1}{c|}{SD} & \multicolumn{1}{c}{MLEA} & \multicolumn{1}{c}{RMSE}  &\multicolumn{1}{c|}{SD} & \multicolumn{1}{c}{MLEA} & \multicolumn{1}{c}{RMSE} & \multicolumn{1}{c|}{SD} \\
\hline
        
100 (53.6\%)	&	-0.4803	&	0.5916	&	0.5943	&	0.7791	&	1.1906	&	1.1963	&	-0.3829	&	0.7859	&	0.7892	&	1.6152	&	2.0228	&	2.0285	\\
250 (52.9\%)	&	-0.4834	&	0.3686	&	0.3701	&	0.6875	&	0.8205	&	0.8222	&	-0.3684	&	0.4272	&	0.4289	&	1.6976	&	0.9705	&	0.9740	\\
500 (51.7\%)	&	-0.5417	&	0.2133	&	0.2102	&	0.8275	&	0.4491	&	0.4445	&	-0.3675	&	0.2499	&	0.2506	&	1.8262	&	0.5080	&	0.5048	\\
1000 (51.3\%)	&	-0.4944	&	0.1770	&	0.1778	&	0.7353	&	0.3550	&	0.3564	&	-0.3734	&	0.1992	&	0.1988	&	1.7966	&	0.3880	&	0.3871	\\
2000 (50.3\%)	&	-0.5096	&	0.1060	&	0.1061	&	0.7872	&	0.2218	&	0.2198	&	-0.3618	&	0.1123	&	0.1122	&	1.7795	&	0.2119	&	0.2108	\\

\end{tabular}\label{LT5}
}
\centering
\smallskip\noindent
\resizebox{\linewidth}{!}{
\begin{tabular}{|c|ccc|ccc|ccc|ccc|}
\toprule
&\multicolumn{6}{c|}{$\overline{\alpha}=1.0$} & \multicolumn{6}{c|}{$\overline{\lambda}=20.085$} \\ 
\cline{2-13}
&\multicolumn{3}{c|}{$\beta_{30}=-1.0$}&\multicolumn{3}{c|}{$\beta_{31}=2.0$}& \multicolumn{3}{c|}{$\beta_{40}=1.25$}&\multicolumn{3}{c|}{$\beta_{41}=3.5$} \\ 
\cline{2-13}
$n$ (CD) & \multicolumn{1}{c}{MLEA} & \multicolumn{1}{c}{RMSE} & \multicolumn{1}{c|}{SD} & \multicolumn{1}{c}{MLEA} & \multicolumn{1}{c}{RMSE} & \multicolumn{1}{c|}{SD} & \multicolumn{1}{c}{MLEA} & \multicolumn{1}{c}{RMSE}  &\multicolumn{1}{c|}{SD} & \multicolumn{1}{c}{MLEA} & \multicolumn{1}{c}{RMSE} & \multicolumn{1}{c|}{SD} \\
\hline
       
100 (53.6\%)	&	-1.0296	&	0.5028	&	0.5044	&	2.3500	&	1.2185	&	1.1730	&	1.1630	&	1.2360	&	1.2392	&	3.6091	&	2.0239	&	2.0311	\\
250 (52.9\%)	&	-1.0163	&	0.2790	&	0.2799	&	2.0894	&	0.5950	&	0.5912	&	1.2754	&	0.5784	&	0.5807	&	3.5367	&	0.9865	&	0.9908	\\
500 (51.7\%)	&	-0.9916	&	0.1567	&	0.1573	&	2.0134	&	0.3377	&	0.3392	&	1.2748	&	0.3872	&	0.3883	&	3.4708	&	0.6564	&	0.6591	\\
1000 (51.3\%)	&	-0.9933	&	0.0956	&	0.0958	&	1.9788	&	0.2247	&	0.2248	&	1.2423	&	0.2144	&	0.2153	&	3.5119	&	0.3202	&	0.3216	\\
2000 (50.3\%)	&	-1.0049	&	0.0731	&	0.0733	&	2.0088	&	0.1516	&	0.1521	&	1.2699	&	0.1571	&	0.1567	&	3.4689	&	0.2240	&	0.2229	\\

\bottomrule
\end{tabular}\label{LT6}
}
\end{table}

\section{Brazilian bank loan portfolio}\label{apl}

In this section we present an application of the proposed model in a database made available by one of the largest Brazilian banks. Our objective is to check if customer characteristics are associated with consumer propensity of being fraudsters, delinquents, or long-term customers, i.e., no chance of becoming delinquent ahead.

It is important to note that the given database, quantities, rates, and levels of the available covariate, do not necessarily represent the actual condition of the financial institution's customer base. Despite being a real database, the bank may have sampled the data in order to change the current status of its loan portfolio.

Figures \ref{sumbase1} and \ref{sumbase2} present a summary of the portfolio together with the portfolio subdivision considering one available covariate. For reasons of confidentiality, we will refer to it by covariate $x$. In this case, we can see that $x$ has three levels, referring to a particular characteristics of a bank's customer profile.

\begin{table}[H]
\centering
\caption{Summary of the bank loan lifetime data.}
\begin{tabular}{|ccccc|}
\toprule
Portfolio 		 & Number of & Number of       & Average of    		    &  Standard deviation of        \\ 
							 & consumers   & defaulters  		  & time to default 		&  time to default    \\ 
\hline
$x=1$          & 1,646  & 305 (18.530\%) &  15.016 					& 11.874 				  \\
$x=2$          & 1,561  & 242 (15.503\%) &  15.641 					& 12.134 				   \\
$x=3$          & 942   & 93  (9.873\%)  &  18.183 					& 12.868 				     \\
\hline
Total					 & 4,149 & 640 (15.425\%) &  15.712 					& 12.148				  \\

\bottomrule
   \end{tabular}\label{sumbase1}
\end{table}

\begin{table}[H]
\centering
\caption{Summary of fraud and censored lifetime data.}
\begin{tabular}{|ccc|}
\toprule
Portfolio 		 & Number of      & Number of   \\ 
							 & fraudsters       & censored   \\ 
\hline
$x=1$          & 157 (9.538\%) & 1184 (71.940\%) \\
$x=2$          & 114 (7.303\%) & 1205 (77.194\%)  \\
$x=3$          &  34 (3.609\%)  & 815 (86.518\%)    \\
\hline
Total 						 & 305 (7.351\%) & 3204 (77.223) \\

\bottomrule
   \end{tabular}\label{sumbase2}
\end{table}

\subsection{Model 1}\label{model1}
                                           
To proceed with the application, we will deal with dummy variables. As $x$ has three levels, then we have two dummy variables, $dx1$ and $dx2$, where $dx1=1$ if $x=1$, and $dx1=0$, otherwise, and similarly, $dx2=1$ if $x=2$, and $dx2=0$, otherwise. The customer group such that $x = 3$ is characterized by setting  $dx1$~$=$~$dx2$~$=0$.
				
According to \ref{aplimod1}, we do not link covariables to the Weibull parameters, which we leave to the next subsection. However, note that we made the following re-parametrization, $\alpha=e^{\alpha^{'}}$ and  $\lambda=e^{\lambda^{'}}$. Thus we have the following set of parameters $\left\{\beta_{10}, \beta_{11}, \beta_{12}, \beta_{20}, \beta_{21}, \beta_{22},\alpha^{'}, \lambda^{'}\right\}$, to be estimated by MLE, as specified in section \ref{model}.

\begin{equation}
 \left\{
\begin{array}{cccc}
(\gamma_{0i},\gamma_{1i}) &=& \left(\frac{e^{\mathbf \beta_{10} + {dx1}_i\beta_{11}+ {dx2}_i\beta_{12}}}{1+e^{\mathbf \beta_{10} + {dx1}_i\beta_{11}+ {dx2}_i\beta_{12}}+e^{\mathbf \beta_{20} + {dx1}_i\beta_{21}+ {dx2}_i\beta_{22}}},\frac{e^{\mathbf \beta_{20} + {dx1}_i\beta_{21}+ {dx2}_i\beta_{22}}}{1+e^{\mathbf \beta_{10} + {dx1}_i\beta_{11}+ {dx2}_i\beta_{12}}+e^{\mathbf \beta_{20} + {dx1}_i\beta_{21}+ {dx2}_i\beta_{22}}}\right),\\
\alpha               &=& e^{\alpha^{'}}, \\
\lambda               &=& e^{\alpha^{'}}, \\
\end{array}
\right.
\label{aplimod1}
\end{equation}

From Table \ref{LT7}, through the analysis of the parameter $\beta_{10}$, we see that being part of group $x=3$ is significant for differentiation on the propensity to fraud, where customers with this characteristic are less likely to be fraudster as customers of other groups.

The parameters $\beta_{20},\beta_{21}$ and $\beta_{22}$, confirm that the segmentation given by the covariable $x$ is significant to establish a descending order of long-term survival rates, from the group $x=3$ with higher cure rate to the lowest cure rate group, $x=1$.

\begin{table}[H]
\centering
\caption{Maximum likelihood estimation results for the zero-inflated Weibull cure rate regression model.}
\begin{tabular}{|cccc|}
\toprule
Parameter& Estimate (est) & Standard error (se) & $|$est$|/$ se  \\ 
\hline
$\beta_{10}$	&	-1.26264	&	0.20315	&	6.21526	\\
$\beta_{11}$	&	0.37604	&	0.22297	&	1.68648	\\
$\beta_{12}$	&	0.27879	&	0.23015	&	1.21133	\\
$\beta_{20}$	&	1.87945	&	0.11966	&	15.70613	\\
$\beta_{21}$	&	-0.81176	&	0.13205	&	6.14738	\\
$\beta_{22}$	&	-0.55874	&	0.13504	&	4.13770	\\
$\alpha^{'}$	&	0.11249	&	0.04326	&	2.60024	\\
$\lambda^{'}$	&	3.16833	&	0.07646	&	41.43616	\\
\bottomrule
   \end{tabular}\label{LT7}
\end{table}

\begin{figure}[H]
	\centering
		\includegraphics[scale=0.25]{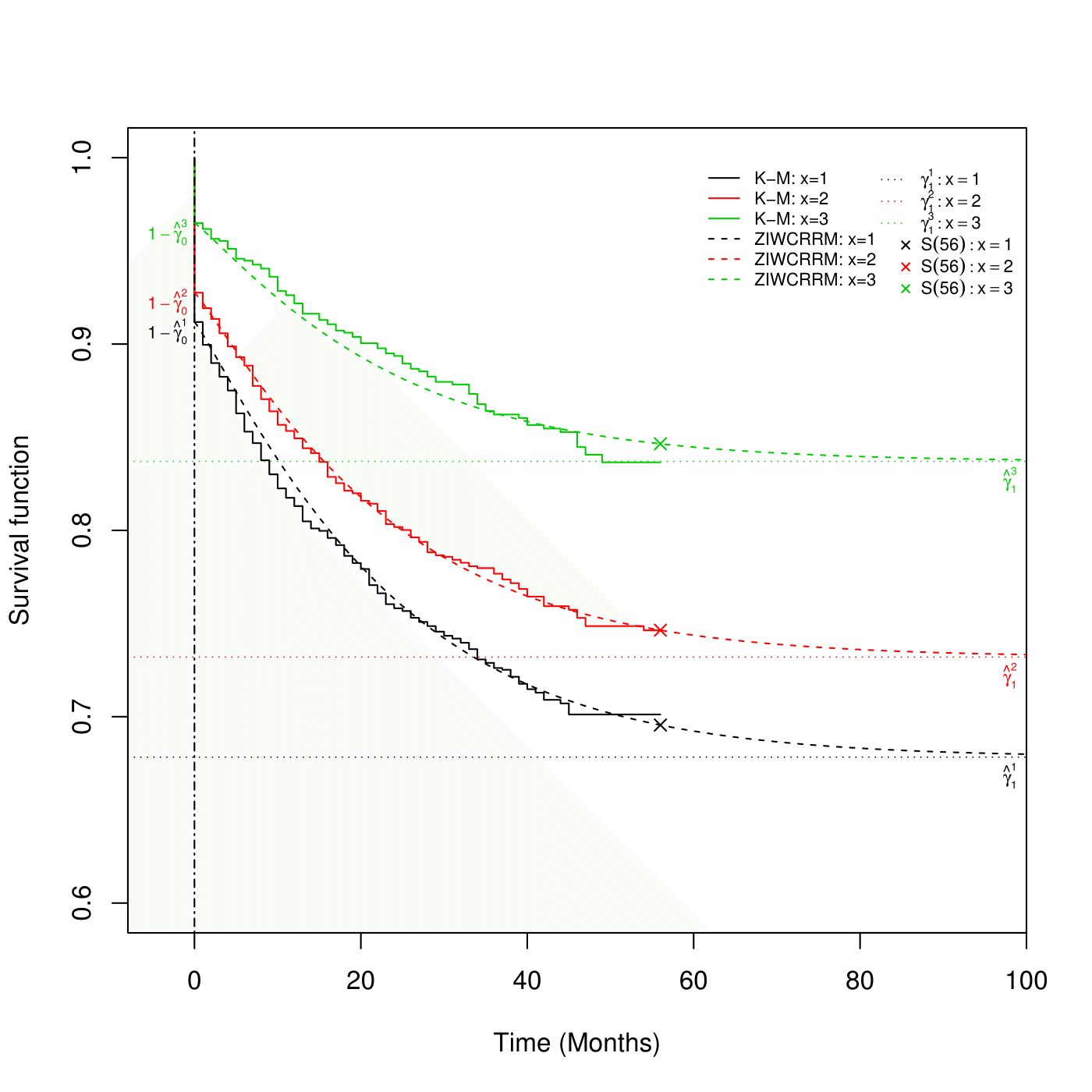}
	\caption{The Kaplan-Meier (K-M) survival curve, according to customer profile $x=1$, $x=2$ and $x=3$.}
	\label{ajuste_LD_ZAW}
\end{figure}

Thus, the estimated parameters ratify the behaviour of groups as shown by the tables \ref{sumbase1} and \ref{sumbase2} and the figure \ref{ajuste_LD_ZAW}.

Table \ref{results1} shows the model results, considering the configuration parameters as given in \ref{aplimod1}. We can see that the outcomes are very satisfactory when compared with the actual data found in tables \ref{sumbase1} and \ref{sumbase2}.

\begin{table}[H]
\centering
\caption{Modelling outcomes of the model 1}
\begin{tabular}{|cccc|}
\toprule
Estimated 		 & Consumer         & Consumer    & Consumer     \\ 
parameter			 & with $x=1$       & with $x=2$  & with $x=3$   \\ 
\hline
$\hat{\gamma}_0$       & 0.0953670        & 0.0730215   & 0.0361181    \\
$\hat{\gamma}_1$       & 0.6731908        & 0.7316650   & 0.8362142    \\
$\hat{S}(56)$          & 0.6902227        & 0.7460382   & 0.8456093    \\
\bottomrule
   \end{tabular}\label{results1}
\end{table}

The Kaplan-Meier (K-M) survival curves for the Brazilian bank loan portfolio, according to customer profiles $x=1$, $x=2$ and $x=3$, are presented in the Figure \ref{ajuste_LD_ZAW}. This figure presents the fitted survival functions as defined by the zero-inflated Weibull cure rate regression (ZIWCRR) model in \ref{zisf}. We observed that the improper survival function falls instantly at $t = 0$ to the values ${1-\hat{\gamma}_0}^1$, ${1-\hat{\gamma}_0}^2$, ${1-\hat{\gamma}_0}^3$, respectively, when $x = 1$, $x = 2$ or $x = 3$. So, the model accommodated well the zero-inflated survival times and, as expected, the plateaus at points ${\hat{\gamma}_1}^1$, ${\hat{\gamma}_1}^2$ and ${\hat{\gamma}_1}^3$ highlighted the presence of cure rate.

Once the maximum observed time up to default for a bank customer is $56$, the values for $S(56)$, represent the estimated survival rate of the sub-portfolio at this point. These values are compared with the cure rates shown in the third column of table \ref{sumbase2}.

\subsection{Model 2}\label{model2}

Finally, we present the entire set of regression parameters that can be related to the covariate $x$. In other words, we also connect the two parameters of the Weibull distribution. Thus, we have the following set of parameters $\left\{\beta_{10}, \beta_{11}, \beta_{12}, \beta_{20}, \beta_{21}, \beta_{22}, \beta_{30}, \beta_{31}, \beta_{32},\beta_{40}, \beta_{41}, \beta_{42},\right\}$ to be estimated by MLE approach.

\begin{equation}
 \left\{
\begin{array}{cccc}
(\gamma_{0i},\gamma_{1i}) &=& \left(\frac{e^{\mathbf \beta_{10} + {dx1}_i\beta_{11}+ {dx2}_i\beta_{12}}}{1+e^{\mathbf \beta_{10} + {dx1}_i\beta_{11}+ {dx2}_i\beta_{12}}+e^{\mathbf \beta_{20} + {dx1}_i\beta_{21}+ {dx2}_i\beta_{22}}},\frac{e^{\mathbf \beta_{20} + {dx1}_i\beta_{21}+ {dx2}_i\beta_{22}}}{1+e^{\mathbf \beta_{10} + {dx1}_i\beta_{11}+ {dx2}_i\beta_{12}}+e^{\mathbf \beta_{20} + {dx1}_i\beta_{21}+ {dx2}_i\beta_{22}}}\right),\\
\alpha_i               &=& e^{\mathbf \beta_{30} + {dx1}_i\beta_{31}+ {dx2}_i\beta_{32}}, \\
\lambda_i               &=& e^{\mathbf \beta_{40} + {dx1}_i\beta_{41}+ {dx2}_i\beta_{42}}, \\
\end{array}
\right.
\label{aplimod2}
\end{equation}

The regression parameters, $\beta_{20}$, $\beta_{30}$ and $\beta_{40}$, reinforce the fact that the group $x = 3$ is an important feature for discriminating customer profiles regarding to the risk assumed in lending.

Unfortunately, in the model 2, most of the estimated parameters were not statistically significant, which may suggest that a higher number of covariates is necessary to better explain the data distribution.

\begin{table}[H]
\centering
\caption{Maximum likelihood estimation results for the zero-inflated Weibull cure rate regression model.}
\begin{tabular}{|lccc|}
\toprule
Parameter& Estimate (est) & Standard error (se) & $|$est$|/$ se  \\ 
\hline
$\beta_{10}$	&		0,1721	&	0,1256	&	1,3702	\\
$\beta_{11}$	&		-0,0825	&	0,1399	&	0,5893	\\
$\beta_{12}$	&		-0,0398	&	0,1432	&	0,2779	\\
$\beta_{20}$	&		3,5164	&	0,2950	&	11,9182	\\
$\beta_{21}$	&		-0,4352	&	0,3129	&	1,3907	\\
$\beta_{22}$	&		-0,3776	&	0,3165	&	1,1933	\\
$\beta_{30}$	&		-1,4281	&	0,2662	&	5,3644	\\
$\beta_{31}$	&		0,5685	&	0,2867	&	1,9829	\\
$\beta_{32}$	&		0,4584	&	0,2934	&	1,5622	\\
$\beta_{40}$	&		1,6870	&	0,2381	&	7,0859	\\
$\beta_{41}$	&		-0,5834	&	0,2535	&	2,3016	\\
$\beta_{42}$	&		-0,3486	&	0,2566	&	1,3582	\\
\bottomrule
   \end{tabular}\label{LT8}
\end{table}

\section{Concluding remarks}\label{con} 

We introduced a methodology based on the zero-inflated cure rate model, in order to assess the propensity to fraud in bank loan applications, which, to the best of our knowledge, has not been applied to account for zero-inflated (fraudster) data in the credit risk environment. 

An advantage of our approach is to accommodate zero-inflated times, which is not possible in the standard cure rate model. In this scenario, information from fraudsters is exploited through the joint modelling of the survival time of fraudsters, who are equal to zero, along with the survival times of the remaining portfolio.

To illustrate the proposed method, a loan survival times of a Brazilian bank loan portfolio was modelled by the proposed zero-inflated Weibull cure rate model, and the results showed satisfactory.

\section*{References}

\bibliography{Reference}

\end{document}